\input{aipcheck}

\documentclass[
    ,final            
  ]
  {aipproc}
\layoutstyle{6x9}

\catcode`\|=11
\let\= = \equiv
\def\_#1{_{\!#1}}
\def\||#1{{\left\Vert#1\right\Vert^{\kern-0.5em\phantom{1}}_{\kern-0.5em\phantom{1}}}}
\def\|#1{{\left\vert#1\right\vert^{\kern-0.5em\phantom{1}}_{\kern-0.5em\phantom{1}}}}
\def\(#1{{\left(#1\right)^{\kern-0.5em\phantom{1}}_{\kern-0.5em\phantom{1}}}}
\def\[#1{{\left[#1\right]^{\kern-0.5em\phantom{1}}_{\kern-0.5em\phantom{1}}}}
\def\{#1{{\left\lbrace#1\rbrace^{\kern-0.5em\phantom{1}}_{\kern-0.5em\phantom{1}}}}

\def\calL{{\cal L}}
\def\cases#1{\def\\{\cr}
          \left\lbrace\>\vcenter{\normalbaselines\m@th
 \ialign{$##\hfil$&&\quad$##\hfil$\crcr#1\crcr}}\right.^{\kern-0.5em\phantom{1}}_{\kern-0.5em\phantom{1}}}

\def\d{{\rm d}}

\def\deldel#1#2{\frac{\delta #1}{\delta #2}}

\def\deq{:\=}

\def\frac#1#2{{#1\over#2}}

\mathchardef\gamma="710D 
\mathchardef\Gamma="7100 

\let\Ga=\Gamma

\mathchardef\lambda="7115 
\mathchardef\Lambda="7103 

\def\mini{{\mu\nu}}

\def\R{{\rm R}}

\def\rbrace{\right\}}

\def\sqr#1#2{{\vcenter{\vbox{\hrule height.#2pt
 \hbox{\vrule width.#2pt height#1pt \kern#1pt
 \vrule width.#2pt}
 \hrule height.#2pt}}}
}
\def\square{\mathchoice\sqr74\sqr74\sqr{2.1}3\sqr{1.5}3\>}

\def\Tag#1{\ifmmode\eqno{(#1)}\hbox to\rightskip{\null}\else(#1)\fi}

\def\ie{{\it i.e.}}

\begin{document}

\title{First Order Extended Gravity and the Dark Side of the Universe: the General Theory}

\classification{95.30.Sf-98.80.Jk}
\keywords      {Dark matter, Dark energy, Dark metric, Gravitation}

\author{S. Capozziello}{
  address={Dipartimento di Scienze Fisiche, Universit\`a di Napoli "Federico II", and
INFN Sez.\ di Napoli, Compl.\ Univ.\ di Monte S.~Angelo, Edificio G, Via Cinthia, I-80126, Napoli, Italy}
}

\author{M. De Laurentis}{
  address={Dipartimento di Scienze Fisiche, Universit\`a di Napoli "Federico II", and
INFN Sez.\ di Napoli, Compl.\ Univ.\ di Monte S.~Angelo, Edificio G, Via Cinthia, I-80126, Napoli, Italy}}

\author{M. Francaviglia}{
  address={Dipartimento di Matematica, Universit\`a di Torino, and INFN Sez.  di Torino,
Via Carlo Alberto 10, 10123 Torino, Italy}
}

\author{S. Mercadante}{
  address={Dipartimento di Matematica, Universit\`a di Torino, and INFN Sez.  di Torino,
Via Carlo Alberto 10, 10123 Torino, Italy}}

\begin{abstract}
 General Relativity is not
the definitive theory of Gravitation due to several shortcomings
which are coming out both from theoretical and experimental
viewpoints. At large scales (astrophysical and cosmological
scales) the attempts to match it with the today observational
data lead to invoke Dark Energy and Dark Matter as the bulk
components of the cosmic fluid. Since no final evidence, at
fundamental level, exists for such ingredients, it is clear that
General Relativity presents shortcomings at infrared scales. On
the other hand, the attempts to formulate theories more general 
than the Einstein one give rise to mathematical difficulties that
need workarounds which, in turn, generate problems from the
interpretative viewpoint.  We present here a completely new
approach to the mathematical objects in terms of which a theory of
Gravitation may be written in a first-order ({\it \`a la\/}
Palatini) formalism, and introduce the concept of {\bf Dark
Metric} which could completely bypass the introduction of
disturbing concepts as Dark Energy and Dark Matter.
\end{abstract}

\maketitle


\section{Introduction}

 Einstein's General Relativity ({\bf GR}) is a self-consistent
theory which dynamically describes space, time and matter under the
same  standard. The result is  a deep and beautiful scheme which,
starting from some first principles, is capable of explaining a
huge number of gravitational phenomena, ranging from laboratory up
to cosmological scales. Its predictions are well tested at  Solar
System scales  and give rise to a comprehensive cosmological
model that agrees with Standard Model of particles, with
recession of galaxies, cosmic nucleosynthesis and so on.
Despite of these good results the recent advent of the so-called
{\it Precision Cosmology} and several tests coming from the Solar
System outskirts (e.g. the Pioneer Anomaly), the self-consistent
scheme of GR seems to disagree with an increasingly high number of
observational data, as {\it e.g.} those coming from IA-type Supernovae,
used as standard candles, large scale structure ranging from
galaxies up to galaxy superclusters and so on. Furthermore, being
not renormalizable, it fails to be quantized in any ``classical''
way (see \cite{1}).  In other words, it seems then, from
ultraviolet up to infrared scales, that GR is not and cannot be
the definitive theory of Gravitation also if it successfully
addresses a wide range of phenomena.
Many attempts have been therefore made  both to recover the
validity of GR at all scales, on the one hand, and to produce
theories that suitably generalize the Einstein one, on the other
hand.
In order to interpret a large number of recent
observational data inside the paradigm of GR, the introduction of
Dark Matter (DM) and Dark Energy (DE) has seemed to be
necessary: the price of preserving the {\sl simplicity\/} of the
Hilbert Lagrangian has been, however, the introduction of rather
odd-behaving physical entities which, up to now, have not been
revealed by any experiment at fundamental scales. In other words
we are observing the large scale effects of missing matter (DM)
and the accelerating behavior of the Hubble flow (DE) but no final
evidence of these ingredients exists, if we want to deal with them
under the standard of  quantum particles or quantum fields. In
Section~3 we shall argue whether, after all, it is really
preferable the use of the {\sl simplest\/} Lagrangian.
An opposite approach resides in the  so-called  Non-Linear
Theories of Gravity (NLTGs), that have been
also investigated by many authors,  in connection with
Scalar-Tensor Theories (STTs).  In this case, no
ill-defined ingredients have to be required, at the price of big
mathematical complications.  None of the many efforts made up to
now to solve this problem (see later) seem to be satisfactory
from an interpretative viewpoint.
What we shortly present here is a completely  new approach to the
mathematical objects in terms of which a theory of Gravitation may
be written, whereby Gravity is encoded from the very beginning in
a (symmetric) linear connection in SpaceTime.  At the end we
shall nevertheless conclude that, although the gravitational field is a linear
connection, the fundamental field of Gravity turns out {\it a
posteriori\/} to be still a metric,  but not the ``obvious'' one
given from the very beginning (which we shall therefore call 
{\it "apparent metric}").  Rather we shall show the relevance of another
metric, ensuing from the gravitational dynamics, that we shall call
{\bf Dark Metric} since we claim it being a possible source of the
{\sl apparently\/} ``Dark Side'' of our Universe which reveals, at
large scales, as missing matter (in clustered structures) and
accelerating behavior (in the Hubble fluid).
To complete our program, we need first to  recall some facts
regarding different (relativistic) theories of Gravitation.  This
will not be an historical {\it compendium}, but just a collection
of speculative hints useful to our aims.

\section{Historical Remarks}

 Einstein devoted  more than ten years (1905--1915/1916) to develop a theory
of Gravitation based on the following requirements (see
\cite{3}): {\bf principle of equivalence} (Gravity and Inertia
are indistinguishable; there exist observers in free fall, {\it i.e.}
inertial motion under gravitational pull); {\bf principle of
relativity} (SR holds pointwise; the structure of SpaceTime is
pointwise Minkowskian); {\bf principle of general covariance}
(``democracy'' in Physics); {\bf principle of causality} (all
physical phenomena propagate respecting the light-cones).
Einstein, who was also deeply influenced by Riemann's teachings
about the link between matter and curvature, decided then to
describe Gravity by means of a (dynamic) SpaceTime $M$ endowed
with a dynamic Lorentzian metric $g$.  This appeared to be a good
choice for a number of reasons: a metric is the right tool to
define measurements (rods \& clocks); the geodesics of a metric
are good mathematical objects to describe the free fall; a
Lorentzian manifold is pointwise Minkowskian; it is suitable to be
the domain of tensor fields; is compatible with a light-cones
structure.  And, after all, at that time, there was no other
geometrical field that Einstein could use to define the curvature of a
differentiable manifold!

Following this approach, Einstein deduced his famous equations:
$$
G_\mini\deq\R_\mini-\frac{1}{2}\>\R(g)\>g_\mini=8\pi\>G\>T_\mini~.\label{1}
$$
A linear concomitant of  the Riemann tensor of $g$, nowadays
called the Einstein tensor $G_\mini$, equals the stress-energy
tensor $T_\mini\deq\deldel{L_{\rm mat}}{g_\mini}$ that reflects
the properties of matter.  Here $\R_\mini$ is the Ricci tensor of
the metric $g$ and $\R(g)\deq g^\mini\>\R_\mini$ is the scalar
curvature of the metric, while $L_{\rm mat}\=\calL_{\rm mat}\>\d
s$ is the matter Lagrangian and $G$ is the coupling constant.
In other words, the distribution of matter influences Gravity
through 10 second-order field equations.  Their structure, in a
sense and {\it mutatis mutandis}, is the same as Newton second law
of Dynamics: no forces means geodesic motion, while the effects of
sources are to produce curvature (just in motion in the Newtonian
case, where the Space and Time are fixed and immutable; both in
the structure of SpaceTime and in its motions in the Einstein
case). GR has been a success:  it admits an elegant and very {\sl
simple\/} Lagrangian formulation ($L_{\rm
H}\deq\R(g)\>\sqrt{g}\>\d s$) and most of its predictions have
been soon experimentally verified and these have remained valid
for many years after its introduction.  So there was no reason for
Einstein to be unhappy with his beautiful creation, at least for
some time. In GR, is $g$ the gravitational field?  Einstein knew  that it is
not, since $g$ is a tensor, while the principle of equivalence
holds true and implies that there exist frames in which the
gravitational field can be inertially switched off, while a tensor
cannot be set to vanish in a frame, if it does not vanish in all
frames.  Free fall is, in fact, described by the geodesics of $(M,
g)$:
\begin{equation}
\label{2}
\ddot{x}^{\lambda}+\Gamma_{\mu\nu}^{\lambda}(g)\dot{x}^{\mu}\dot{x}^{\nu}=0
\end{equation}
and Einstein himself argued that  the right objects to represent
the gravitational field have to be the Christoffel symbols $\Gamma_{\mu\nu}^{\lambda}(g)$; 
the metric $g$ is just the potential of
the gravitational field,  but being Christoffel symbols
algorithmically constructed out of $g$, the metric remains the
fundamental variable: $g$ gives rise to the gravitational field,
to causality, to the principle of equivalence as well as to rods
and  clocks.
In 1917, working on the  theory of ``parallelism'' in manifolds,
T. Levi-Civita understood that parallelism and curvature are
non-metric features of space, but rather features of ``affine''
type, having to do with ``congruences of privileged lines'' (see
\cite{4}).  Generalizing the case of Christoffel symbols
$\Gamma^{\lambda}_{\mu\nu}(g)$ of a metric $g$, Levi-Civita introduced
the notion of {\bf linear connection} as the most general object
$\Gamma^{\lambda}_{\mu\nu}$ such that the equation of geodesics (autoparallel curves, in fact) 

\begin{equation}\label{3}
\ddot x^\lambda+\Gamma^\lambda_{\mu\nu}\>\dot x^\mu\>\dot x^\nu=0
\end{equation}
is generally covariant.  This revolutionary idea  (that stands in
fact at the heart of Non-Euclidean Geometries) was
immediately captured by Einstein, who, unfortunately, did not
further use it up to its real strength.  We shall come back later
on this topic, as this work is strongly based on it.
Even if it was clear to Einstein that Gravity  induces ``freely
falling observers'' and that the principle of equivalence selects,
in fact, an object that cannot be a tensor, since it is capable of
being ``switched off'' and set to vanish at least in a point, he
was obliged to choose it under the form of the linear connection
$\Gamma_{\rm LC}(g)$, given locally by Christoffel symbols of $g$, now called the Levi-Civita connection (of $g$),
 fully determined by the metric structure
itself. Einstein, for obvious reasons, was very satisfied of
having reduced all SpaceTime structure and Gravity into a single
geometrical object.
Still, in 1918, H. Weyl tried (see \cite{5}) to unify
Gravity with Electromagnetism, using for the first time a linear
connection defined over SpaceTime, assumed as a dynamical field
non-trivially depending on a metric.  Weyl's idea failed because
of a wrong choice of the Lagrangian and few more issues, but it
generated however a keypoint: connections may have a physically
interesting dynamics.
Einstein  soon showed a great interest in Weyl's idea.  He too
began to play with connections, in the obsessed seek for the
``geometrically'' Unified Theory.  But he never arrived to
``dethronize'' $g$ in the description of the gravitational field.
Probably, in some moments, he was not so happy with the fact that
the gravitational field is not the fundamental object, but just a
by-product of the metric; however, he never really changed his
mind about the role of $g$.
In 1925 Einstein constructed a theory  that depends on a metric
$g$ and a symmetric linear connection $\Gamma$, to be varied
independently (the so-called, because of a misunderstanding with
W.\ Pauli, {\bf Palatini method}; see \cite{6}); he defined in fact a
Lagrangian theory in which the gravitational Lagrangian is
\begin{equation}
\label{4}
L_{\rm PE}: \equiv R(g,\Gamma)\>\sqrt{g}\>d s,
\end{equation}
There are now 10 + 40 independent variables  and the field
equations, in vacuum, are:

\begin{eqnarray}
 R_{(\mu\nu)}-\frac{1}{2}R(g,\Gamma)g_{\mu\nu}=0\nonumber\\
 \nabla^{\Gamma}_{\alpha}(\sqrt{g}g^{\mu\nu})=0\nonumber\\
\label{6}
\end{eqnarray}
where $R_{(\mu\nu)}$ is the symmetric part of the Ricci tensor of $\Gamma$
$R_{\mu\nu}(\Gamma,\partial\Gamma)$ and $\nabla^{\Gamma}$ denotes
covariant derivative with respect to $\Gamma$.

Since the dimension of spacetime is greater than $2$, the second
field equation \eqref{6}$_{2}$ constrains the connection $\Gamma$, which
is {\it a priori\/} arbitrary, to coincide {\it a posteriori} with
the Levi-Civita connection of the metric $g$ (Levi-Civita
theorem).  By substituting this information into the first field
equation \eqref{6}$_1$, the vacuum Einstein equation for $g$ is
eventually obtained.  In Palatini formalism, the metric $g$
determines a priori rods \& clocks, while the connection $\Gamma$ the free
fall, but since {\it a posteriori\/} the same result of GR is
found, Einstein soon ceased to show a real interest in this formalism.
 The situation does not change if matter
is present through a matter Lagrangian $L_{\rm mat}$ (independent
of $\Gamma$ but just depending on $g$ and other external matter
fields), that generates an energy-momentum tensor $T_{\mu\nu}$ as
$T_{\mu\nu}:\equiv \frac{\delta{L_{\rm mat}}}{\delta{g_{\mu\nu}}}$.
  If the total Lagrangian is then assumed to be $L_{\rm tot}:\equiv L_{\rm
PE}+L_{\rm mat}$ field equation \eqref{6}$_{1}$ are reflected by

\begin{equation}
\label{7}
R_{\mu\nu}-\frac{1}{2}R(g,\Gamma)g_{\mu\nu}=8\pi G T_{\mu\nu}
\end{equation}
and again \eqref{6}$_2$ implies,  {\it a posteriori}, that \eqref{7}
reduces eventually to Einstein equations in presence of matter.

Let us also emphasize that  the dynamical coincidence between
$\Gamma$ and the Levi-Civita connection of $g$ is entirely due to the
particular Lagrangian considered by Einstein, which is the {\sl
simplest},  but not the only possible one!  Furthermore, it seems
to us that Einstein did not fully recognize that the Palatini
method privileges  the affine structure with respect to the metric
structure. Notice that, in this case (i.e. in Palatini formalism), the
relations
\begin{equation}
\Gamma^{\lambda}_{\mu\nu}=\Gamma^{\lambda}_{\mu\nu}(g)
\label{8}
\end{equation}
are field equations: the fact that $\Gamma$ is the Levi-Civita
connection of $g$ is no longer an assumption {\it a priori\/} but
it is the dynamical outcome of field equations!

\section{The geometric structure of gravitational theories }

Let us begin this Section with a digression of fundamental importance in our schema.
Thanks to the work of Levi-Civita and Einstein we know that every manifold having to do with gravity may be endowed
with at least two a priori distinct structures:
\itemize{
\item A {\it Riemannian metric structure}, defined by a (Lorentzian) metric tensor. It is responsible for
the definition of (pseudo)distances and angles. It selects, on the space on which is defined,
a class of hyper-surfaces as the level sets of the distance function (the analogous
of circles in the ordinary Plane Geometry) and a class of curves, called  the {\it  geodesics of the metric}, 
or  {$g$-geodesics} in this report, defined
as the stationary (minimum) length paths connecting pairs of points.
\item An {\it affine structure} defined by an affine connection. It is responsible for the parallel
transport along curves through the definition of the notion of covariant derivation.
It selects, on the space on which is defined, the notion of straightness of
lines, i.e. a class of self-parallel lines (curves whose tangent vector is parallel transported
by the connection, the analogous of straight lines in the ordinary
Plane Geometry) called the {\it geodesics of the connection}, or {\it $\Gamma$-geodesics} in this
work. As it is well known, on any such curve the notion of arc length is defined
(independently of a metric).}

 It is crucial to understand that, since the metric and the affine structures
are a priori independent, the {\it $g$-geodesics} and the {\it $\Gamma$-geodesics} are, in principle very
different. In exactly the same way, an arc length of a {\it $\Gamma$-geodesics} may not coincide  with the distance of its extrema. 
This abstraction process is quite difficult to be performed,
since the Geometry we are used to does not need it. Nevertheless it is a great
opportunity, as we shall show in the sequel. As mathematical physicists we should
have seen this kind of picture more than once....!
In ordinary Euclidean Geometry, the metric and the affine structures (which together
correspond to the well-known compass \& unmarked straightedge Geometry)
are, actually, deeply intertwined. A strong link between these two structures is in fact
set by the {\it simplest} variational principle: straight lines are the shortest path between
any two points. This is exactly to say: {\it $\Gamma$-geodesics} coincide with the {\it $g$-geodesics}.
Formulated in the latter way, this situation is not peculiar of the ordinary Euclidean
Geometry. On the contrary, it applies to all Euclidean and Non-Euclidean Geometries.
This is why we have to force our {\it intuitus} to accept the more general case of
geometries where {\it $\Gamma$-geodesics}  do differ  from the {\it $g$-geodesics}.
With these considerations in mind, let us reconsider the Palatini method we introduced
in the previous Section. We believe it is clear enough that it can be considered
an attempt of Einstein to force his {\it intuitus}. Not the first of his life. But this attempt,
so to say, failed. We have already observed that the reason of this
failure is entirely due to the {\it simplicity} of the Hilbert Lagrangian. This is why we
regard to more general theories as an opportunity, not only as a complication.
Before to delve into these more general theories, let us remark a final point about
the metric and the affine structures. From a purely geometrical viewpoint, they stand
on an equal footing. But from a physical point of view the situation is different.
The fundamental principle of Newtonian Physics (the First Law, {\it i.e.} the Principle of
Inertia) selects, in fact, the straight lines as the more fundamental structure: in absence
of forces motions are rectilinear and uniform, {\it i.e.} 4-dimensionally straight. Circles
instead limit their role to the definition of space distances. In GR something similar
happens with the Principle of Equivalence (which in some sense is the generalization
of the Newtonian Principle of Inertia): it selects the geodesics, the {\it $\Gamma$-geodesics}, as
the most important lines from a gravitational point of view.

\section{Extended Theories of Gravity}

All that said, we believe we should first seriously reconsider
NLTGs, without being unsensitive with respect to the appeal of
{\sl simplicity}, in the spirit of Occam Razor. This is why we
begin to restrict ourselves to the first level of generalization
of GR, the so-called {\bf $f(R)$-theories} of metric type (see
{ e.g.}\
\cite{7} for a  review of the results concerning
these theories). Here $f$ denotes any ``reasonable'' function of
one-real variable. The Lagrangian is assumed to be
\begin{equation}
L_{\rm NL}(g):\equiv f(R(g))\sqrt{g}d s
\label{9}
\end{equation}
where $R(g)=g^{\mu\nu}R_{\mu\nu}(g)$ is the scalar curvature of $(M,g)$.

Of course, from $f(R)$-theories, we know that GR is retrieved in,
and only in, the particular case $f(R)\equiv R$, i.e. if and only if
the Lagrangian is linear\footnote{Of corse if $f(R) = R + \Lambda$ we have the Einstein Equations with a cosmological constant $\Lambda$.} in $R$.

Let us recall here just a few  keypoints on metric
$f(R)$-theories.

When treated in  the purely metric formalism, these theories are
mathematically much more complicated than GR. These theories do in
fact produce field equations that are of the fourth order in the
metric:

\begin{equation}
f'(R(g))R_{\mu\nu}(g)-\frac{1}{2}f(R(g))g_{\mu\nu}
  -\underbrace{(\nabla_{\mu}\nabla_{\nu}-g_{\mu\nu}\square)\>f'(R(g))}_{\hbox{$4^{\rm th}$ order term}}
  =8\pi\>G\>T\_{\mu\nu}
\label{10}
\end{equation}
where $f'$ denotes the derivative of $f$ with  respect to its real
argument.  This is something that cannot be accepted if one
believes that physical laws should be governed by second order
equations. In \eqref{10} we see a second order part that resembles
Einstein tensor (and reduces identically to it if and only if
$f(R)=R$, i.e. if and only if $f'(R)=1$) and a fourth order
``curvature term'' (that again reduces to zero if and only if
$f(R)=R$).

A first workaround that was suggested long ago  to this problem is
to push the 4th order part $(\nabla_{\mu}\nabla_{\nu}-g_{\mu\nu}\square)\>f'(R(g))$ 
to the r.h.s.  of these equations. This
lets us to interpret it as an ``extra gravitational stress''
$T^{~{\rm curv}}_{\mu\nu}$ due to higher-order curvature effects,
much in the spirit of Riemann.  In any case, however, the fourth
order character of these equations makes them  unsuitable under
several aspects, so that they were eventually abandoned for long
time and only recently they have regained  interest (see
\cite{7} and references therein).

A second way to tackle the problem  has been proposed in 1987 (see
\cite{8}, based on earlier work by the same authors
\cite{9}, together with the references quoted therein). Notice
that these are the first papers where the Legendre transformation
that introduces an extra scalar field has been ever considered in
literature (it has been later ``re-discovered'' by other authors),
so that its priority should always be appropriately quoted when
dealing with ``metric'' $f(R)$-theories. This is a method {\it
\`a la\/} Hamilton, in which, whenever one has a non-linear
gravitational metric Lagrangian of the most general type 
$L_{\rm GNL}(g):\equiv f(g,Ric(g))$, one defines a {\sl second\/} metric $p$
as\footnote {This idea corresponds to an Einstein's attempt, dating back to 1925, to construct a ``purely
       affine'' theory  (see \cite{10}), i.e. a theory in which the only dynamical field is a linear
       connection.  In this theory no metric is given from the beginning, but since it is obviously
       necessary to have a metric, the problem arise of how to construct it out of a connection.
       Einstein first tried to define the metric as the symmetric part of the Ricci tensor
       constructed out of the connection.  But this idea could not work (unless for quadratic
       Lagrangians). A.Eddington then
       proposed the recipe \eqref{11}. In this way Einstein and Eddington obtained a theory that
       reproduces GR,
       without introducing anything new. That is why Einstein eventually abandoned it too. On these purely  affine
       theories see also \cite{11},  where J. Kijowski correctly pointed out that in the purely affine
       framework the prescription \eqref{11} of Einstein and Eddington is nothing but the assumption
       that the metric can be considered as a momentum canonically conjugated to the connection.
       Of corse $p$ is a true metric if it is nondegenerate, something that 
       is always true in an open set of the space of all solutions and for a ÒgenericÓ function $f$.}

\begin{equation}
p_{\mu\nu}:\equiv \frac{\partial L_{\rm grav}}{\partial R_{\mu\nu}}~.
\label{11}
\end{equation}
In this way  the second metric $p$, in fact a non-degenerate metric for $f$ nowhere vanishing,
a canonically conjugated
momentum for $g$, is a function of $g$ together with its first and
second derivatives, since it is a function of $g$ and $Ric(g)$,
the Ricci tensor of $g$.  Notice that this leads to two equations
of the second order in $g$ and $p$, as Hamilton method always
halves the order of the equations by doubling the variables.
Following this method in the simpler $f(R)$ case one gets that
the ``auxiliary'' metric $p$ is related to the original one $g$ by
a conformal transformation:

\begin{equation}
p\equiv\phi g,\qquad \phi:\equiv f'(R(g)).
\label{12}
\end{equation}
The Lagrangian equations \eqref{10}  are then rewritable as a
Hamiltonian system:
\begin{equation}
 {\rm Ein}(p)=T_{\rm mat}+T_{\rm KGnl}
 \end{equation}
 \begin{equation}
 {\rm KGnl}(\phi)=0
\label{13}
\end{equation}
where KGnl  means non-linear Klein-Gordon (because of a potential
depending on $f$; see \cite{7},
\cite{8}, \cite{9} for details).
Rewritten in this form, the theory has now two variables: the
``auxiliary'' metric $p$ (or the original one $g$) and the scalar field $\phi$.  This is why
these theories are called {\bf Scalar-Tensor Theories}.  For more
details, and in particular for their application in Cosmology and
Extragalactic Astrophysics, see, {\it e.g.},
\cite{7} and the references quoted therein.
Notice that \cite{8}, \cite{9} and all subsequent literature
left in fact open a few fundamental problems: {\it Who really are the
second metric $p$ and the scalar field $\phi$? How to interpret
them (the scalar field $\phi$ survives even in vacuum)?  And\dots
what about the original metric $g$?}

Fortunately there is a third method to solve the problem.

\section{The Palatini Approach and the  Dark Metric}

The third method  anticipated at the end of the previous Section
is the Palatini method applied to the case of $f(R)$-theories.
Now SpaceTime is no longer a couple $(M,g)$ but rather a triple
$(M,g,\Gamma)$, with $\Gamma$ symmetric for simplicity and convenience.
 The Lagrangian is assumed to be
the non-linear Palatini-Einstein Lagrangian

\begin{equation}
L_{\rm NLPE}(g,\Gamma):\equiv f(R(g,\Gamma))\>\sqrt{g}d s
\label{14}
\end{equation}
with $R(g,\Gamma):\equiv g^{\mu\nu}R_{\mu\nu}(\Gamma,\partial\Gamma)$ and $f$
``reasonable.'' Field equations \eqref{6} are now replaced by the
following:

\begin{equation}
f'(R(g,\Gamma))R_{(\mu\nu)}-\frac{1}{2}\>f(R(g,\Gamma))\>g_{\mu\nu}=G\>T_{\mu\nu}
 \end{equation}
 \begin{equation}
 \nabla^{\Gamma}_\alpha(f'(R(g,\Gamma))\sqrt{g}\>g^{\mu\nu})=0
\label{15}
\end{equation}
that take into account a possible  Lagrangian of the type $L_{\rm
mat}=L_{\rm mat}(g,\psi)$, with $\psi$ arbitrary matter fields coupled
to $g$ alone (and {\sl not\/} to $\Gamma$).  Notice that \eqref{15}$_1$
reduces to \eqref{7} if and only if $f(R)\equiv R$.  Notice also that
the trace of equation \eqref{15}$_1$ gives
\begin{equation}
R(g,\Gamma)\>f'(R(g,\Gamma))-\frac{m}{2}\>f(R(g,\Gamma))=G\tau
\label{16}
\end{equation}
being $\tau:\equiv g^{\mu\nu}\>T_{\mu\nu}$ the ``trace''  of the
energy-momentum tensor. This equation has been called the {\bf master
equation}  \cite{13} and it is  at the basis of a subtle
discussion of ``universality'' of Einstein equations in non-linear
special cases (i.e., when $\tau\equiv0$). Notice also the analogy of
\eqref{16} with the trace of \eqref{10}, i.e.
\begin{equation}
f'(R(g))R(g)-\frac{m}{2}\>f(R(g))+(m-1)\>\square f'(R(g))=G\tau
\label{17}
\end{equation}
and notice that only in the peculiar  case $f(R)=R$ they reduce
to the {\sl same\/} equation, namely \eqref{7}. In all other cases
\eqref{17} entails that non-linearity ($f'\not=1$) produces, in the
metric formalism, effects due to the scalar factor $f'(R)$, i.e.
depending eventually on a scalar field tuned up by the curvature of $\Gamma$.
Approaching $f(R)$-theories {\it \`a la\/}  Palatini, we may now
follow \cite{13} step-by-step and make a number of
considerations (well summarized also in the recent critical review
\cite{14}).  At the end of these considerations we may conclude
that:
\begin{enumerate}
 \item When (and only when) $f(R(g,\Gamma))=R(g,\Gamma)$ then  GR is ``fully'' recovered for the given
  metric $g$.
 \item For a generic $f(R(g,\Gamma))$, in presence of matter such that $g^{\mu\nu}\>T_{\mu\nu}=0$ (and thence,
  in particular, in vacuum), the theory
  is still equivalent to GR for the given metric $g$ with a ``quantized'' cosmological constant
  $\Lambda$ and a modified coupling constant. In this case, in fact, the master equation
  \eqref{16}  implies that the scalar curvature $R(g,\Gamma)$ has to be a suitable constant, possibly
  and usually not unique but always chosen in a set that depends on $f$, so that \eqref{15}$_2$ still
  implies \eqref{8} with an additional cosmological term $\Lambda g_{\mu\nu}$.
 \item For a generic $f(R(g,\Gamma))$ but in presence of matter such that
  $g^{\mu\nu}T_{\mu\nu}\not=0$ one can
  solve the master equation \eqref{16}, for $m\not=2$,  and obtain $R(g,\Gamma)$ as a function
  $R(\tau)$ of the given trace $\tau$. Then, knowing $f$, one gets implicitly
  $f(R(g,\Gamma))=f(\tau)$ and $f'(R(g,\Gamma))=f'(\tau)$, so that equation \eqref{15}$_2$ tells us that
  $\Gamma$ is forced to be the Levi-Civita connection $\Gamma_{\rm LC}(h)$ of a new metric $h$,
  conformally related to the original one $g$ by the relation
 \begin{equation}
  h_{\mu\nu}:\equiv f'(\tau)\>g_{\mu\nu}=f'(R(g,\Gamma))g_{\mu\nu}~.
  \label{18}
  \end{equation}
  Then, using again equation \eqref{15}$_1$, we see  that the theory could be still rewritable as in
  \eqref{12} in a purely metric setting, but with far less interpretative problems, as we can
  immediately show. From a viewpoint ``{\it \`a la  Palatini}'' in a genuine sense the method
  has in fact generated a completely new perspective.  The remaining field equations \eqref{15}$_1$,
  in fact, are still equivalent to Einstein equations with matter (and cosmological constant)
  provided one changes the metric from $g$ to $h$!
  \end{enumerate}
The most enlightening case is that of $f(R)$ with generic matter
and $\tau\not=0$.  Here, in fact, the universality property (see
again \cite{13}) does not hold in his strict form, but in an
interesting wider interpretation: the dynamics of the connection
$\Gamma$ still forces $\Gamma$ itself to be the Levi-Civita connection
of a metric, but not of the ``original'' metric $g$, which we
shall prefer to call the {\bf apparent metric} for a reason we
clarify in a moment.  Instead, the dynamics of $\Gamma$ identifies a
new metric $h$, conformally related to the apparent one $g$, which
we call the {\bf Dark Metric}. The Dark Metric $h$, we claim, could well be
the true origin of the ``Dark Side of the Universe''!
The apparent metric $g$ is in fact the one by means of which we
perform measurements in our local laboratories.  In other words,
the metric $g$ is the one we have to use every day to construct
and read instruments (rods \& clocks).  This is why we like to
call it the ``apparent'' metric.  But we claim that the right
metric we have to use as the  fundamental object to describe
Gravity is, by obvious reasons, the Dark Metric, since it is the
one responsible for gravitational free fall through the
identification $\Gamma=\Gamma_{\rm LC}(h)$. Notice, incidentally, that
photon world-lines and causality are not changed, since the
light-cones structure of $g$ and $h$ are the same by conformal
invariance.
In other words, in our laboratories we have to use the apparent
metric $g$, but in our Gravity theories the dark one $h$.  The
translation from one ``language'' to the other is nothing but the
conformal factor $f'(R)=f'(\tau)$, which manifestly depends on the theory
and on its content in ordinary matter.  Let us also notice
explicitly that this in particular implies that if a certain
metric $h$ is expected to be a solution of a problem, from a
theoretical point of view, it is rather important to look for $h$
in experiments.  Testing our theories with $g$, in a sense, is
wrong, since it is the  conformally related metric $h$ to be
searched for instead when dealing with Gravitation!

\section{Concluding Remarks}

 The (unknown)
conformal factor $\phi\=f'(\R(g,\Ga))\=f'(\tau)$ has to be
phenomenologically tested against observational data in order to
find which is the (class of) Lagrangian(s) $f$ that, given $\tau$
(\ie, given the ``visible matter''), allow one to interpret the
``supposed Dark Matter'' (and Energy) as a curvature effect
\cite{15}. 
In  the other contribution by the same authors in these Proceedings, we will give hints on how DE (accelerated
cosmic behavior) and DM (clustered structures) could be
interpreted as Dark Metric effects according to the lines dveloped here.

\end{document}